\documentclass[pra,twocolumn,showpacs,showkeys]{revtex4-1}

\usepackage{graphicx}
\usepackage{amsmath}
\usepackage{amssymb}
\usepackage{mathrsfs}
\usepackage{siunitx}
\usepackage{pstricks}
\usepackage{psfrag}
\usepackage{xspace}
\usepackage{color}
\usepackage{upgreek}
\usepackage[english]{babel}
\usepackage[normalem]{ulem}
\usepackage{filemod}
\usepackage{hyperref}

\hypersetup{colorlinks=true,breaklinks=true,citecolor={blue},linkcolor={blue},urlcolor={blue},pdfstartview=FitH}

\newcommand{\NoAutoSpaceBeforeFDP}{}
\newcommand{\AutoSpaceBeforeFDP}{}

\newcommand{\ie}{{\it i.e.}\@\xspace}
\newcommand{\eg}{{\it e.g.}\@\xspace}

\newcommand{\eq}[1]{Eq.~\eqref{#1}}
\newcommand{\eqs}[1]{Eqs.~\eqref{#1}}

\newcommand{\fig}[1]{Fig.~\ref{#1}}

\renewcommand{\bm}[1]{\boldsymbol{\mathbf{#1}}}
\newcommand{\ud}{\mathrm{d}}
\newcommand{\bra}{\left\langle}
\newcommand{\ket}{\right\rangle}

\newcommand{\im}{\operatorname{Im}}
\newcommand{\re}{\operatorname{Re}}
\newcommand{\He}{\operatorname{\Theta}}

\newcounter{tempa}
\newcounter{tempb}
\newcounter{tempc}
\newcounter{tempd}
\newenvironment{diaga}[1]{\psset{unit=1.5mm,fillstyle=solid,fillcolor=white}
   \begin{pspicture}[shift=-0.15](0,-1)(#1,3)}{\end{pspicture}
}
\newenvironment{diagb}[1]{\psset{unit=1.5mm,fillstyle=solid,fillcolor=white}
   \begin{pspicture}[shift=-0.15](0,-1)(#1,6)}{\end{pspicture}
}
\newenvironment{diagc}[1]{\psset{unit=1.5mm,fillstyle=solid,fillcolor=white}
   \begin{pspicture}[shift=-0.15](0,-1)(#1,2)}{\end{pspicture}
}
\newcommand{\ginc}[2]{\psline(#1,0)(#2,0)}
\newcommand{\einc}[2]{\psline[linestyle=dashed](#1,0)(#2,0)}
\newcommand{\gmoy}[2]{\psline[linewidth=0.5](#1,0)(#2,0)}

\newcommand{\particule}[1]{\pscircle(#1,0){1}}

\newcommand{\correldeux}[2]{
   \setcounter{tempa}{#2}
   \addtocounter{tempa}{-#1}
   \divide \value{tempa} by 2
   \setcounter{tempb}{#1}
   \addtocounter{tempb}{#2}
   \divide \value{tempb} by 2
   \psarc[linestyle=dashed,fillstyle=none](\value{tempb},0){\value{tempa}}{0}{180}
}
\newcommand{\correltrois}[3]{
   \setcounter{tempa}{#1}
   \addtocounter{tempa}{#3}
   \divide \value{tempa} by 2
   \psline[linestyle=dashed](\value{tempa},6)(#1,0)
   \psline[linestyle=dashed](\value{tempa},6)(#2,0)
   \psline[linestyle=dashed](\value{tempa},6)(#3,0)
}

\newenvironment{ddiag}[1]{\psset{unit=1.5mm,fillstyle=solid,fillcolor=white}
   \begin{pspicture}[shift=-3.15](0,-4)(#1,4)}{\end{pspicture}
}
\newenvironment{ddiaga}[1]{\psset{unit=1.5mm,fillstyle=solid,fillcolor=white}
   \begin{pspicture}[shift=-3.15](0,-4)(#1,6)}{\end{pspicture}
}

\newenvironment{ddiagd}[1]{\psset{unit=1.5mm,fillstyle=solid,fillcolor=white}
   \begin{pspicture}[shift=-3.15](0,-4)(#1,9)}{\end{pspicture}
}
\newenvironment{ddiage}[1]{\psset{unit=1.5mm,fillstyle=solid,fillcolor=white}
   \begin{pspicture}[shift=-8.15](0,-9)(#1,4)}{\end{pspicture}
}
\newenvironment{ddiagf}[1]{\psset{unit=1.5mm,fillstyle=solid,fillcolor=white}
   \begin{pspicture}[shift=-8.15](0,-9)(#1,9)}{\end{pspicture}
}
\newcommand{\gginc}[3]{\psline(#1,#3)(#2,#3)}
\newcommand{\eeinc}[3]{\psline[linestyle=dashed](#1,#3)(#2,#3)}
\newcommand{\ggmoy}[3]{\psline[linewidth=0.5](#1,#3)(#2,#3)}

\newcommand{\pparticule}[2]{\pscircle(#1,#2){1}}

\newcommand{\ccorreldeuxa}[2]{
   \setcounter{tempa}{#2}
   \addtocounter{tempa}{-#1}
   \divide \value{tempa} by 2
   \setcounter{tempb}{#1}
   \addtocounter{tempb}{#2}
   \divide \value{tempb} by 2
   \psarc[linestyle=dashed,fillstyle=none](\value{tempb},3){\value{tempa}}{0}{180}
}
\newcommand{\ccorreldeuxb}[2]{
   \setcounter{tempa}{#2}
   \addtocounter{tempa}{-#1}
   \divide \value{tempa} by 2
   \setcounter{tempb}{#1}
   \addtocounter{tempb}{#2}
   \divide \value{tempb} by 2
   \psarc[linestyle=dashed,fillstyle=none](\value{tempb},-3){\value{tempa}}{180}{360}
}
\newcommand{\ccorreldeuxc}[4]{
   \psline[linestyle=dashed](#1,#2)(#3,#4)
}

\begin{document}

\title{Absorption of scalar waves in correlated disordered media and its maximization using stealth hyperuniformity}

\author{A. Sheremet}
\affiliation{ESPCI Paris, PSL University, CNRS, Institut Langevin, 1 rue Jussieu, F-75005, Paris, France}
\author{R. Pierrat}
\affiliation{ESPCI Paris, PSL University, CNRS, Institut Langevin, 1 rue Jussieu, F-75005, Paris, France}
\email{romain.pierrat@espci.psl.eu}
\author{R. Carminati}
\affiliation{ESPCI Paris, PSL University, CNRS, Institut Langevin, 1 rue Jussieu, F-75005, Paris, France}
\email{remi.carminati@espci.psl.eu}

\date{\filemodprint{\jobname}~~File: \jobname}

\begin{abstract}
   We develop a multiple scattering theory for the absorption of waves in disordered media.  Based on a general
   expression of the average absorbed power, we discuss the possibility to maximize absorption by using structural
   correlations of disorder as a degree of freedom. In a model system made of absorbing scatterers in a transparent
   background, we show that a stealth hyperuniform distribution of the scatterers allows the average absorbed power to
   reach its maximum value.  This study provides a theoretical framework for the design of efficient non-resonant
   absorbers made of dilute disordered materials, for broadband and omnidirectional light, and other kinds of waves.
\end{abstract}

\maketitle 

\section{Introduction}\label{sec:introduction}

Multiple scattering of light and other kinds of waves in disordered media has been extensively studied in the last
decades.  The field has been initially driven by fundamental questions in mescoscopic
physics~\cite{SHENG-2006,MONTAMBAUX-2007}, and by applications in sensing and imaging~\cite{SEBBAH-2001,RIPOLL-2012}. It
has been recognized that spatial correlations in the disorder strongly influence the properties of a scattering
medium~\cite{MAURICE-1957,MARET-1990,CAO-2003-1,SCHEFFOLD-2004,SEGEV-2011,CONLEY-2014}, and the concept of correlated
materials --a class of materials being neither fully random nor perfectly crystalline-- has
emerged~\cite{CEFE-LOPEZ-2010,VYNCK-2012,BURRESI-2014,FROUFE-PEREZ-2017}.  Engineering disorder may become a novel
approach in the design of photonic materials with specific
functionnalities~\cite{MULLER-2013,LESEUR-2016-1,SELLERS-2017,MOLESKY-2018,MILOSEVIC-2019}.

An interesting feature of correlated disorder is the possibility to tune the level of absorption. Enhancing absorption
with disordered materials, while keeping other properties (\eg electrical conduction or mechanical response) unaffected,
is a key issue for light harvesting in photovoltaic
devices~\cite{SARGENT-2012,GALVEZ-2014,MUPPARAPU-2015,KOMAN-2016,LEE-2017}. The possibility of using spatial
correlations to enhance substantially the level of absorption in disordered materials has been highlighted
recently~\cite{DAL_NEGRO-2012,VYNCK-2012,LESEUR-2016,WANG-2018,LIU-2018,FLORESCU-2018}.  Upper bounds for absorbance
enhancement have been derived, which provide constrains that can guide optimization
processes~\cite{HUGONIN-2015,MILLER-2016,BIGOURDAN-2019}.  It is also interesting to recall that enhanced absorption
motivated early studies of Anderson localization of light~\cite{JOHN-1984}. It is interesting to note that
absorption enhancement is also possible through the concept of coherent perfect absorption, which requires an action on
a coherent incident wavefront~\cite{CAO-2010-1,CAO-2011,CHONG-2011,HSU-2015,LIEW-2016}.

In a recent contribution, some of us have shown that in a model material made of discrete absorbing scatterers in a
transparent matrix, the average absorbed power can be maximized by distributing the scatterers on a disordered stealth
hyperuniform point pattern~\cite{BIGOURDAN-2019}. This absorption enhancement in non-crystalline (partially ordered)
distributions of scatterers was shown to be broadband with a wide angular acceptance (contrary to coherent resonant
absorption that is usually limited to narrow frequency and angular ranges). The main goal of the present paper is to
develop a theoretical framework to analyze absorption in correlated disordered media, and to prove the relevance of
stealth hyperuniform patterns in maximizing the average absorbed power. The paper is organized as follows. In
Sec.~\ref{sec:avg_abs_power} we derive an exact (non-perturbative) expression for the average power absorbed in a
disordered medium using multiple scattering theory. In Sec.~\ref{sec:asymptotic_RTE}, we show that in the
weak-extinction regime, the expression of the absorbed power is consistent with radiative transfer theory in the
appropriate large-scale limit. In this framework, we prove in Sec.~\ref{sec:max} that, in a medium made of discrete
absorbing scatterers in a transparent background, the average absorbed power can be maximized when the scatterers are
distributed on a stealth hyperuniform pattern. Interestingly, we also prove that although spatial correlations
substantially influence the scattering mean free path, they actually leave the absorption mean free path unaffected in
the weak-extinction regime. Finally we briefly summarize the main results in Sec.~\ref{sect:conclusion}.

\section{Average absorbed power}\label{sec:avg_abs_power}

We consider a disordered medium described by a position and frequency dependent dielectric function of the form
$\epsilon(\bm{r},\omega)=1+\delta\epsilon(\bm{r},\omega)$, where $\delta\epsilon$ is a random variable assumed to be
complex valued. The imaginary part of $\delta\epsilon$ accounts for absorption in the medium. We further assume
that $\delta\epsilon$ has a correlation function
\begin{equation} \label{eq:correlation_epsilon}
   \bra \delta\epsilon(\bm{r},\omega) \delta\epsilon^*(\bm{r}^\prime,\omega) \ket = 
   \bra \delta\epsilon(\bm{r},\omega) \ket \bra \delta\epsilon^*(\bm{r}^\prime,\omega) \ket + C(\bm{r}-\bm{r}^\prime),
\end{equation}
where $\langle \ldots \rangle$ denotes statistical averaging over realizations of disorder, and $*$ stands for complex
conjugate. Here we assume that the medium is statistically homogeneous and isotropic, with $C(\bm{r}-\bm{r}^\prime)$
depending only on $|\bm{r}-\bm{r}^\prime|$. The correlation function above implies that the real and imaginary
parts of the dielectric function are perfectly correlated. This general permittivity model includes the particular case
of absorbing particles distributed in a transparent matrix, which will be considered in
Sec.~\ref{subsec:min_p_s_wilde}. Also note that we do not require $\delta\epsilon$ to have zero mean, since its
imaginary part is non-negative in a passive absorbing medium.  Upon illumination by an external optical field, and in
the scalar wave model that we use in this study, the local average absorbed power per unit volume is~\cite{JACKSON-1962}
\begin{equation}\label{eq:avg_abs_power}
   \bra \mathcal{P}_a(\bm{r},\omega)\ket=\frac{1}{2}\re\bra j(\bm{r},\omega)E^*(\bm{r},\omega)\ket,
\end{equation}
where $E(\bm{r},\omega)$ is the electric field and $j(\bm{r},\omega)$ the current density. Since
$j(\bm{r},\omega)=-i\omega\epsilon_0 \,\delta\epsilon(\bm{r},\omega)E(\bm{r},\omega)$, we immediately find that 
\begin{equation}\label{eq:avg_abs_power2}
   \bra \mathcal{P}_a(\bm{r},\omega)\ket=\frac{\epsilon_0\omega}{2}
      \im \bra\delta\epsilon(\bm{r},\omega)E(\bm{r},\omega)E^*(\bm{r},\omega)\ket
\end{equation}
where $\epsilon_0$ is the vacuum permittivity. We shall now derive a general expression for $\bra
\mathcal{P}_a(\bm{r},\omega)\ket$ in the framework of the standard perturbative theory of multiple scattering.

\subsection{Electric field and multiple scattering}

We start by introducing a diagrammatic expansion of the electric field. For the sake of simplicity, we omit the
dependence on $\omega$, keeping in mind that all fields are monochromatic. The electric field obeys the Helmholtz
equation
\begin{equation}\label{eq:wave_equation}
   \Delta E(\bm{r})+k_0^2\epsilon(\bm{r})E(\bm{r})=S(\bm{r})
\end{equation}
where $k_0=\omega/c=2\pi/\lambda$ is the vacuum wavenumber, $c$ being the speed of light and $\lambda$ the wavelength in
free space, and $S(\bm{r})$ is an external source. Likewise, the incident field $E_0$ is the solution to
\begin{equation}\label{eq:wave_equation_inc}
   \Delta E_0(\bm{r})+k_0^2E_0(\bm{r})=S(\bm{r}).
\end{equation}
From \eqs{eq:wave_equation} and (\ref{eq:wave_equation_inc}), it is easy to see that the scattered field $E_s=E-E_0$ obeys
\begin{equation}\label{eq:wave_equation_sca}
   \Delta E_s(\bm{r})+k_0^2E_s(\bm{r})=-k_0^2\delta\epsilon(\bm{r})E(\bm{r}) .
\end{equation}
The free-space Green function $G_0$ is the solution to
\begin{equation}\label{eq:wave_equation_green}
   \Delta G_0(\bm{r},\bm{r}_0)+k_0^2G_0(\bm{r},\bm{r}_0)=-\delta(\bm{r}-\bm{r}_0)
\end{equation}
with an outgoing wave condition when $|\bm{r}-\bm{r}_0| \to \infty$, and is given by
\begin{equation}\label{eq:green_function_expr}
   G_0(\bm{r},\bm{r}_0)= \dfrac{\exp(ik_0|\bm{r}-\bm{r}_0|)}{4\pi|\bm{r}-\bm{r}_0|}.
\end{equation}
Using \eqs{eq:wave_equation_sca} and \eq{eq:wave_equation_green}, the scattered field can be written as
\begin{equation}\label{eq:lippmann-schwinger}
   E_s(\bm{r})= \int G_0(\bm{r},\bm{r}')V(\bm{r}')E(\bm{r}')\ud^3r',
\end{equation}
where $V(\bm{r}')=k_0^2\delta\epsilon(\bm{r}')$ is the scattering potential. The total field obeys the integral equation
\begin{equation}\label{eq:lippmann-schwinger}
   E(\bm{r})=E_0(\bm{r})+\int G_0(\bm{r},\bm{r}')V(\bm{r}')E(\bm{r}')\ud^3r',
\end{equation}
which is known as the Lippmann-Schwinger equation.
It can be rewritten formally using operator notation, in the form
\begin{equation}\label{eq:lippmann-schwinger_formal}
   E=E_0+G_0VE.
\end{equation}
Upon iterating this equation, we obtain the Born series
\begin{equation}\label{eq:cluster}
   E=E_0+G_0VE_0+G_0VG_0VE_0+G_0VG_0VG_0VE_0+\ldots
\end{equation}
that can be understood as a multiple scattering expansion. For practical calculations, it is useful to write the above
expansion using diagrams~\cite{FRISCH-1967}, with the following rules: A circle denotes a scattering event by the
potential $V$, a solid line represents a free-space Green function $G_0$, and a dotted line stands for the incident
field $E_0$. Following these rules, \eq{eq:cluster} becomes
\begin{equation}\label{eq:cluster_diag}
   E=
   \begin{diagc}{8}
      \einc{1}{7}
   \end{diagc}
   +
   \begin{diagc}{14}
      \ginc{1}{7}
      \einc{7}{13}
      \particule{7}
   \end{diagc}
   +
   \begin{diagc}{20}
      \ginc{1}{7}
      \ginc{7}{13}
      \einc{13}{19}
      \particule{7}
      \particule{13}
   \end{diagc}
   +
   \ldots
\end{equation}
The same expansion can be written for the complex conjugate $E^*$. 

\subsection{Diagramatic calculation of average absorbed power}

We now turn to the diagrammatic expansion of $VEE^*$, which is at the root of the computation of the average absorbed
power, as can be seen from \eq{eq:avg_abs_power2}. Using the upper line for $VE$ and the bottom line for $E^*$, we can
write
\begin{widetext}
\begin{multline}\label{eq:cluster_quad_diag}
   VEE^*=
   \begin{ddiag}{8}
      \eeinc{1}{7}{3}
      \eeinc{1}{7}{-3}
      \pparticule{1}{3}
   \end{ddiag}
   +
   \begin{ddiag}{14}
      \eeinc{1}{7}{3}
      \gginc{1}{7}{-3}
      \eeinc{7}{13}{-3}
      \pparticule{1}{3}
      \pparticule{7}{-3}
   \end{ddiag}
   +
   \begin{ddiag}{14}
      \gginc{1}{7}{3}
      \eeinc{7}{13}{3}
      \eeinc{7}{13}{-3}
      \pparticule{1}{3}
      \pparticule{7}{3}
   \end{ddiag}
   +
   \begin{ddiag}{14}
      \gginc{1}{7}{3}
      \gginc{1}{7}{-3}
      \eeinc{7}{13}{3}
      \eeinc{7}{13}{-3}
      \pparticule{1}{3}
      \pparticule{7}{-3}
      \pparticule{7}{3}
   \end{ddiag}
   +
   \begin{ddiag}{20}
      \eeinc{1}{7}{3}
      \gginc{1}{7}{-3}
      \gginc{7}{13}{-3}
      \eeinc{13}{19}{-3}
      \pparticule{1}{3}
      \pparticule{7}{-3}
      \pparticule{13}{-3}
   \end{ddiag}
\\
   +
   \begin{ddiag}{20}
      \gginc{1}{7}{3}
      \gginc{7}{13}{3}
      \eeinc{13}{19}{3}
      \eeinc{1}{7}{-3}
      \pparticule{1}{3}
      \pparticule{7}{3}
      \pparticule{13}{3}
   \end{ddiag}
   +
   \begin{ddiag}{20}
      \gginc{1}{7}{3}
      \eeinc{7}{13}{3}
      \gginc{1}{7}{-3}
      \gginc{7}{13}{-3}
      \eeinc{13}{19}{-3}
      \pparticule{1}{3}
      \pparticule{7}{3}
      \pparticule{7}{-3}
      \pparticule{13}{-3}
   \end{ddiag}
   +
   \begin{ddiag}{20}
      \gginc{1}{7}{3}
      \gginc{7}{13}{3}
      \eeinc{13}{19}{3}
      \gginc{1}{7}{-3}
      \eeinc{7}{13}{-3}
      \pparticule{1}{3}
      \pparticule{7}{3}
      \pparticule{13}{3}
      \pparticule{7}{-3}
   \end{ddiag}
   +
   \begin{ddiag}{20}
      \gginc{1}{7}{3}
      \gginc{7}{13}{3}
      \eeinc{13}{19}{3}
      \gginc{1}{7}{-3}
      \gginc{7}{13}{-3}
      \eeinc{13}{19}{-3}
      \pparticule{1}{3}
      \pparticule{7}{3}
      \pparticule{13}{3}
      \pparticule{7}{-3}
      \pparticule{13}{-3}
   \end{ddiag}
   +\ldots
\end{multline}
\end{widetext}
where the expansion has been limited to third-order scattering. Note that the two lines only differ by the presence of a 
scattering event $V$ on the left of each diagram in the upper line. Next, we proceed with the calculation of the ensemble average
of \eq{eq:cluster_quad_diag} over all possible configurations of disorder. We start by considering two consecutive scattering events, 
\ie $VG_0V$. We can write
\begin{equation}\label{eq:correlations}
   \bra VG_0V\ket=\bra V\ket G_0\bra V\ket+\bra VG_0V\ket_c ,
\end{equation}
where the second term accounts for structural correlations in the disorder, and is often referred to as the connected part (\ie the part that cannot be factorized).
In terms of diagrams, the above equation is
\begin{equation}\label{eq:correlations_diag}
   \bra VG_0V\ket=
   \begin{diagc}{8}
      \ginc{1}{7}
      \particule{1}
      \particule{7}
   \end{diagc}
   +
   \begin{diagb}{8}
      \ginc{1}{7}
      \correldeux{1}{7}
      \particule{1}
      \particule{7}
   \end{diagb}.
\end{equation}
Here the dotted link in the second diagram means that the two scattering events are spatially correlated.
The presence of correlations affects the higher-order terms in a similar way.  For example, for three scattering events, we have
\begin{multline}\label{eq:correlations_diag_three}
   \bra VG_0VG_0V\ket=
   \begin{diagc}{14}
      \ginc{1}{7}
      \ginc{7}{13}
      \particule{1}
      \particule{7}
      \particule{13}
   \end{diagc}
   +
   \begin{diagb}{14}
      \ginc{1}{7}
      \ginc{7}{13}
      \correldeux{1}{7}
      \particule{1}
      \particule{7}
      \particule{13}
   \end{diagb}
\\
   +
   \begin{diagb}{14}
      \ginc{1}{7}
      \ginc{7}{13}
      \correldeux{7}{13}
      \particule{1}
      \particule{7}
      \particule{13}
   \end{diagb}
   +
   \begin{diagb}{14}
      \ginc{1}{7}
      \ginc{7}{13}
      \correldeux{1}{13}
      \particule{1}
      \particule{7}
      \particule{13}
   \end{diagb}
   +
   \begin{diagb}{14}
      \ginc{1}{7}
      \ginc{7}{13}
      \correltrois{1}{7}{13}
      \particule{1}
      \particule{7}
      \particule{13}
   \end{diagb}.
\end{multline}
We now make use of this averaging method to calculate $\bra VEE^* \ket$, and consider separately two situations.

In the first situation, we assume that the first scattering event in \eq{eq:cluster_quad_diag}, \ie the upper left
circle in each diagram in the upper line, is not correlated to any other scattering event, or is connected to a
scattering event in the upper line only. The expansion below shows typical diagrams contributing to $\bra VEE^*\ket$ in
this situation:
\begin{multline}\label{eq:avg_cluster_quad_1_diag}
   \bra VEE^*\ket_1=
   \begin{ddiag}{20}
      \gginc{1}{7}{3}
      \gginc{7}{13}{3}
      \eeinc{13}{19}{3}
      \gginc{1}{7}{-3}
      \eeinc{7}{13}{-3}
      \pparticule{1}{3}
      \pparticule{7}{3}
      \pparticule{13}{3}
      \pparticule{7}{-3}
   \end{ddiag}
\\
   +
   \begin{ddiaga}{20}
      \gginc{1}{7}{3}
      \gginc{7}{13}{3}
      \eeinc{13}{19}{3}
      \gginc{1}{7}{-3}
      \gginc{7}{13}{-3}
      \eeinc{13}{19}{-3}
      \ccorreldeuxa{1}{7}
      \pparticule{1}{3}
      \pparticule{7}{3}
      \pparticule{13}{3}
      \pparticule{7}{-3}
      \pparticule{13}{-3}
   \end{ddiaga}
   +
   \begin{ddiaga}{20}
      \gginc{1}{7}{3}
      \gginc{7}{13}{3}
      \eeinc{13}{19}{3}
      \gginc{1}{7}{-3}
      \gginc{7}{13}{-3}
      \eeinc{13}{19}{-3}
      \ccorreldeuxa{1}{7}
      \ccorreldeuxc{13}{3}{13}{-3}
      \pparticule{1}{3}
      \pparticule{7}{3}
      \pparticule{13}{3}
      \pparticule{7}{-3}
      \pparticule{13}{-3}
   \end{ddiaga}
   +\ldots
\end{multline}
Next we introduce the self-energy $\Sigma$, defined as 
\begin{equation}\label{eq:sigma_diag}
   \Sigma=
   \begin{diagc}{2}
      \particule{1}
   \end{diagc}
   +
   \begin{diagb}{8}
      \ginc{1}{7}
      \correldeux{1}{7}
      \particule{1}
      \particule{7}
   \end{diagb}
   +
   \begin{diagb}{14}
      \ginc{1}{7}
      \ginc{7}{13}
      \correldeux{1}{13}
      \particule{1}
      \particule{7}
      \particule{13}
   \end{diagb}
   +\ldots
\end{equation}
The self-energy describes the propagation of the average field $\bra E\ket$, and is a central quantity in multiple scattering theory~\cite{SHENG-2006,MONTAMBAUX-2007,RYTOV-1989}. Indeed, by averaging \eq{eq:cluster_diag}, we obtain
\begin{multline}\label{eq:avg_field_diag}
   \bra E\ket=
   \begin{diagc}{8}
      \einc{1}{7}
   \end{diagc}
   +
   \begin{diagc}{14}
      \ginc{1}{7}
      \einc{7}{13}
      \particule{7}
   \end{diagc}
   +
   \begin{diagc}{20}
      \ginc{1}{7}
      \ginc{7}{13}
      \einc{13}{19}
      \particule{7}
      \particule{13}
   \end{diagc}
\\
   +
   \begin{diaga}{20}
      \ginc{1}{7}
      \ginc{7}{13}
      \einc{13}{19}
      \correldeux{7}{13}
      \particule{7}
      \particule{13}
   \end{diaga}
   +
   \ldots
\end{multline}
Using \eq{eq:sigma_diag}, it can be seen that the average field obeys
\begin{equation} \label{eq:dyson_field_op}
   \bra E \ket = E_0 + G_0 \Sigma \bra E \ket ,
\end{equation}
which is known as the Dyson equation. Likewise, the average Green function obeys
\begin{equation} \label{eq:dyson_Green_op}
   \bra G \ket = G_0 + G_0 \Sigma \bra G \ket .
\end{equation}
In real space, the Dyson equation for the electric field is
\begin{multline}\label{eq:dyson_field}
   \bra E(\bm{r})\ket=E_0(\bm{r})
      +\int G_0(\bm{r},\bm{r}')\Sigma(\bm{r},\bm{r}')\bra E(\bm{r}')\ket\ud^3r'.
\end{multline}
It is also useful to introduce the diagramatic expansion of the field correlation function $\bra E E^* \ket$, which
takes the following form:
\begin{multline} \label{eq:diag_EE}
   \bra EE^* \ket = 
   \begin{ddiag}{8}
      \eeinc{1}{7}{3}
      \eeinc{1}{7}{-3}
   \end{ddiag}
   +
   \begin{ddiag}{14}
      \gginc{1}{7}{3}
      \eeinc{7}{13}{3}
      \gginc{1}{7}{-3}
      \eeinc{7}{13}{-3}
      \pparticule{7}{3}
      \pparticule{7}{-3}
   \end{ddiag}
   +
   \begin{ddiag}{14}
      \gginc{1}{7}{3}
      \eeinc{7}{13}{3}
      \gginc{1}{7}{-3}
      \eeinc{7}{13}{-3}
      \ccorreldeuxc{7}{3}{7}{-3}
      \pparticule{7}{3}
      \pparticule{7}{-3}
   \end{ddiag}
\\
   +\ldots
\end{multline}
From \eqs{eq:avg_cluster_quad_1_diag}, (\ref{eq:sigma_diag}) and (\ref{eq:diag_EE}), it can be seen that
\begin{equation} \label{eq:VEE_Sigma}
\bra VEE^*\ket_1= \Sigma \bra E E^* \ket ,
\end{equation}
since all ``horizontal'' correlations in the upper line are included in $\Sigma$, while all ``vertical'' correlations, as in the third diagram in \eq{eq:avg_cluster_quad_1_diag}, are
included in $\bra EE^*\ket$. Finally, using \eqs{eq:avg_abs_power2} and (\ref{eq:VEE_Sigma}), we obtain a first contribution to the
average absorbed power density
\begin{multline}\label{eq:avg_abs_power_1}
   \bra\mathcal{P}_a(\bm{r})\ket_1=
\\
      \frac{\epsilon_0 c^2}{2\omega}\im\left[\int\Sigma(\bm{r},\bm{r}')
         \bra E(\bm{r}')E^*(\bm{r})\ket\ud^3r'\right].
\end{multline}

In the second situation, we consider that the first scattering event in \eq{eq:cluster_quad_diag} is correlated to one or several scattering
events in the lower line, and in addition potentially correlated to one or several events in the upper line 
(\ie we consider all diagrams not taken into account in the previous situation).
In this situation, the expansion of $\bra VEE^*\ket$ takes the form
\begin{multline}\label{eq:avg_cluster_quad_2_diag}
   \bra VEE^*\ket_2=
   \begin{ddiag}{20}
      \gginc{1}{7}{3}
      \gginc{7}{13}{3}
      \eeinc{13}{19}{3}
      \gginc{1}{7}{-3}
      \eeinc{7}{13}{-3}
      \ccorreldeuxc{1}{3}{7}{-3}
      \pparticule{1}{3}
      \pparticule{7}{3}
      \pparticule{13}{3}
      \pparticule{7}{-3}
   \end{ddiag}
   +
   \begin{ddiaga}{20}
      \gginc{1}{7}{3}
      \gginc{7}{13}{3}
      \eeinc{13}{19}{3}
      \gginc{1}{7}{-3}
      \gginc{7}{13}{-3}
      \eeinc{13}{19}{-3}
      \ccorreldeuxc{1}{3}{13}{-3}
      \pparticule{1}{3}
      \pparticule{7}{3}
      \pparticule{13}{3}
      \pparticule{7}{-3}
      \pparticule{13}{-3}
   \end{ddiaga}
\\
   +
   \begin{ddiagd}{20}
      \gginc{1}{7}{3}
      \gginc{7}{13}{3}
      \eeinc{13}{19}{3}
      \gginc{1}{7}{-3}
      \gginc{7}{13}{-3}
      \eeinc{13}{19}{-3}
      \ccorreldeuxa{1}{13}
      \ccorreldeuxc{7}{3}{7}{-3}
      \pparticule{1}{3}
      \pparticule{7}{3}
      \pparticule{13}{3}
      \pparticule{7}{-3}
      \pparticule{13}{-3}
   \end{ddiagd}
   +
   \begin{ddiaga}{20}
      \gginc{1}{7}{3}
      \gginc{7}{13}{3}
      \eeinc{13}{19}{3}
      \gginc{1}{7}{-3}
      \gginc{7}{13}{-3}
      \eeinc{13}{19}{-3}
      \ccorreldeuxc{1}{3}{7}{-3}
      \ccorreldeuxc{13}{3}{13}{-3}
      \pparticule{1}{3}
      \pparticule{7}{3}
      \pparticule{13}{3}
      \pparticule{7}{-3}
      \pparticule{13}{-3}
   \end{ddiaga}
   +\ldots
\end{multline}
in which a few typical diagrams have been represented.
In order to simplify this expansion, we introduce the intensity vertex $\Gamma$, defined by
\begin{equation}\label{eq:gamma_diag}
   \Gamma=
   \begin{ddiag}{2}
      \ccorreldeuxc{1}{3}{1}{-3}
      \pparticule{1}{3}
      \pparticule{1}{-3}
   \end{ddiag}
   +
   \begin{ddiagd}{14}
      \ccorreldeuxa{1}{13}
      \ccorreldeuxc{7}{-3}{7}{3}
      \gginc{1}{7}{3}
      \gginc{7}{13}{3}
      \pparticule{1}{3}
      \pparticule{7}{3}
      \pparticule{13}{3}
      \pparticule{7}{-3}
   \end{ddiagd}
   +
   \begin{ddiage}{14}
      \ccorreldeuxb{1}{13}
      \ccorreldeuxc{7}{-3}{7}{3}
      \gginc{1}{7}{-3}
      \gginc{7}{13}{-3}
      \pparticule{1}{-3}
      \pparticule{7}{-3}
      \pparticule{13}{-3}
      \pparticule{7}{3}
   \end{ddiage}
   +\ldots
\end{equation}
The vertex $\Gamma$ is another important quantity in mutliple scattering theory, and describes the behavior
of the field correlation function $\bra EE^*\ket$~\cite{SHENG-2006,MONTAMBAUX-2007,RYTOV-1989}. Indeed, 
using \eqs{eq:sigma_diag}, (\ref{eq:dyson_Green_op}), (\ref{eq:diag_EE}) and (\ref{eq:gamma_diag}), it can be shown
that the field correlation function obeys
\begin{equation} \label{eq:bethe_salpeter_op}
\bra EE^* \ket = \bra E \ket \bra E^* \ket + \bra G \ket \bra G^* \ket \Gamma \bra E E^* \ket ,
\end{equation}
which is known as the Bethe-Salpeter equation. In real space, this equations reads as
\begin{multline}\label{eq:bethe-salpeter}
   \bra E(\bm{r})E^*(\bm{\uprho})\ket
      =\bra E(\bm{r})\ket\bra E^*(\bm{\uprho})\ket
\\
      +\int \bra G(\bm{r},\bm{r}')\ket\bra G^*(\bm{\uprho},\bm{\uprho}')\ket
         \Gamma(\bm{r}',\bm{\uprho}',\bm{r}'',\bm{\uprho}'')
\\\times
         \bra E(\bm{r}'')E^*(\bm{\uprho}'')\ket
         \ud^3r'\ud^3r''\ud^3\rho'\ud^3\rho''.
\end{multline}
Using \eqs{eq:sigma_diag}, (\ref{eq:dyson_Green_op}), (\ref{eq:avg_cluster_quad_2_diag}) and (\ref{eq:gamma_diag}),
it can be seen that
\begin{equation} \label{eq:VEE_Gamma}
\bra VEE^*\ket_2= \bra G^* \ket \Gamma \bra E E^* \ket,
\end{equation}
which, after insertion into \eq{eq:avg_abs_power2}, leads to the second contribution to the average absorbed power density:
\begin{multline}\label{eq:avg_abs_power_2}
   \bra\mathcal{P}_a(\bm{r})\ket_2=
      \frac{\epsilon_0 c^2}{2\omega}\im\left[\int\bra G^*(\bm{r},\bm{\uprho})\ket
      \right.
\\\left.\vphantom{\int}\times
         \Gamma(\bm{r},\bm{\uprho},\bm{r}',\bm{\uprho}')
         \bra E(\bm{r}')E^*(\bm{\uprho}')\ket\ud^3r'\ud^3\rho\ud^3\rho'\right].
\end{multline}

In summary, we have shown that the average absorbed power density can be written as
\begin{equation}\label{eq:avg_abs_power}
   \bra\mathcal{P}_a(\bm{r})\ket=\bra\mathcal{P}_a(\bm{r})\ket_1
      +\bra\mathcal{P}_a(\bm{r})\ket_2,
\end{equation}
where the two terms on the right-hand side are given by \eqs{eq:avg_abs_power_1} and \eqref{eq:avg_abs_power_2}, respectively.
This result has been derived for scalar waves without any approximation using the diagrammatic description of multiple scattering.
In a medium with finite size occupying a volume $\mathcal{V}$, the total absorbed power $\bra P_a \ket$ is obtained by computing
the integral $\bra P_a \ket = \int_{\mathcal{V}} \bra\mathcal{P}_a(\bm{r})\ket \ud^3r$.

The fact that the expression of $\bra\mathcal{P}_a\ket$ only involves the self-energy $\Sigma$ and intensity vertex
$\Gamma$ makes the result easy to generalize beyond the model of continuous disorder. This can be done by writing
$\Sigma$ and $\Gamma$ in terms of the T-matrix of the medium, which itself can be written in terms of T-matrices of
individual scatterers~\cite{VAN_ROSSUM-1997}.  It is important to note that \eq{eq:avg_abs_power} is
not self-contained. Indeed, computing the integrals in \eqs{eq:avg_abs_power_1} and \eqref{eq:avg_abs_power_2} requires
the knowledge of $\bra G\ket$ and $\bra EE^*\ket$, that are solutions to the Dyson and Bethe-Salpeter equations,
respectively. 

In the following, it will prove useful to rewrite \eq{eq:avg_abs_power} in Fourier space. Assuming statistical
translationnal invariance, the Fourier transforms of the self-energy, average Green function and intensity vertex take
the form
\begin{align}\label{eq:th_reduced_sigma}
   \Sigma(\bm{k},\bm{k}') & =2\pi\delta(\bm{k}-\bm{k}')\widetilde{\Sigma}(\bm{k}),
\\\label{eq:th_reduced_avg_green}
   \bra G(\bm{k},\bm{k}')\ket & =2\pi\delta(\bm{k}-\bm{k}')\bra\widetilde{G}(\bm{k})\ket,
\\\label{eq:th_reduced_gamma}
   \begin{split}
      \Gamma(\bm{k},\bm{k}',\bm{\upkappa},\bm{\upkappa}') &
         =8\pi^3\delta(\bm{k}-\bm{k}'-\bm{\upkappa}+\bm{\upkappa}')
   \\
      & \qquad\times\widetilde{\Gamma}(\bm{k},\bm{k}',\bm{\upkappa},\bm{\upkappa}').
   \end{split}
\end{align}
Using the above expressions, \eq{eq:avg_abs_power} can be rewitten as
\begin{widetext}
   \begin{multline}\label{eq:avg_abs_power_fourier}
      \bra\mathcal{P}_a(\bm{r})\ket=
         \frac{\epsilon_0 c^2}{2\omega}\im\left[
            \int \widetilde{\Sigma}\left(\bm{k}+\frac{\bm{q}}{2}\right)
               \bra E\left(\bm{k}+\frac{\bm{q}}{2}\right)E^*\left(\bm{k}-\frac{\bm{q}}{2}\right)\ket
               \exp(i\bm{q}\cdot\bm{r})\frac{\ud^3k}{8\pi^3}\frac{\ud^3q}{8\pi^3}
         \right.
   \\\left.
            +\int \bra\widetilde{G}^*\left(\bm{k}-\frac{\bm{q}}{2}\right)\ket
               \widetilde{\Gamma}\left(
                  \bm{k}+\frac{\bm{q}}{2},\bm{k}-\frac{\bm{q}}{2},\bm{k}'+\frac{\bm{q}}{2},\bm{k}'-\frac{\bm{q}}{2}
               \right)
               \bra E\left(\bm{k}'+\frac{\bm{q}}{2}\right)E^*\left(\bm{k}'-\frac{\bm{q}}{2}\right)\ket
               \exp(i\bm{q}\cdot\bm{r})\frac{\ud^3k}{8\pi^3}\frac{\ud^3k'}{8\pi^3}\frac{\ud^3q}{8\pi^3}
         \right].
   \end{multline}
\end{widetext}
In absence of absorption, the right-hand side in \eq{eq:avg_abs_power_fourier} has to vanish. It is interesting to note
that this condition is in agreement with the Ward identity, as shown in App.~\ref{app:Ward}. Consistency with the
Ward identity, or equivalently with energy conservation, is an important feature of the multiple scattering theory
developed in this section. Expression (\ref{eq:avg_abs_power_fourier}) is the starting point of the derivation of an
asymptotic form of the average absorbed power in Sec.~\ref{sec:asymptotic_RTE}, and of the discussion of the
maximization of absorption using structural correlations as a degree of freedom in Sec.~\ref{sec:max}.

\section{Weak extinction and radiative transfer limit}\label{sec:asymptotic_RTE}

In this section, we derive an asymptotic expression of the average absorbed power density in the weak-extinction
regime, and show that, in this limit, the usual picture of radiative transfer theory is recovered. The condition of weak
extinction can be written as $k_0\ell_e\gg 1$, where $\ell_e$ is the extinction mean free path.
Qualitatively, $\ell_e$ is the average distance between successive extinction events (scattering or
absorption). It will be defined precisely below [see \eq{eq:extinction_length}]. In this regime, the average field
$\bra E\ket$ and average intensity $\bra I\ket=\bra |E|^2\ket$ vary on the scale of the mean free path, which is much
larger than the wavelength. This means that a large-scale approximation can be performed in
\eq{eq:avg_abs_power_fourier}, that is, we can assume $|\bm{q}|\ll\{|\bm{k}|,|\bm{k}'|\}$. Under this approximation, it
is known that the average intensity obeys the Radiative Transfer Equation (RTE). The detailed derivation of the RTE is
beyond the scope of this paper, and can be found for example in Refs.~\onlinecite{RYTOV-1989,APRESYAN-1996,CAZE-2015}.
Here we briefly summarize the main steps.  We start by Fourier transforming the Bethe-Sapeter equation
(\ref{eq:bethe-salpeter}) in Fourier space, which leads to
\begin{widetext}
   \begin{multline}\label{eq:bethe-salpeter_fourier}
      \bra E\left(\bm{k}+\frac{\bm{q}}{2}\right)E^*\left(\bm{k}-\frac{\bm{q}}{2}\right)\ket
         = \bra \widetilde{G}\left(\bm{k}+\frac{\bm{q}}{2}\right)\ket\bra\widetilde{G}^*\left(\bm{k}-\frac{\bm{q}}{2}\right)\ket
   \\ \times
      \int \widetilde{\Gamma}\left(\bm{k}+\frac{\bm{q}}{2},\bm{k}-\frac{\bm{q}}{2},\bm{k}'+\frac{\bm{q}}{2},\bm{k}'-\frac{\bm{q}}{2}\right)
         \bra E\left(\bm{k}'+\frac{\bm{q}}{2}\right)E^*\left(\bm{k}'-\frac{\bm{q}}{2}\right)\ket
         \frac{\ud^3k'}{8\pi^3}.
   \end{multline}
\end{widetext}
In this equation we have neglected the exponentially small contribution of the average field on the right-hand side,
since we assume that the intensity is calculated far from the external source $S$.  The Dyson equation
(\ref{eq:dyson_Green_op}) can also be written as
\begin{equation}\label{eq:dyson_fourier}
   \bra \widetilde{G}(\bm{k})\ket=G_0(\bm{k})+G_0(\bm{k})\widetilde{\Sigma}(\bm{k})\bra\widetilde{G}(\bm{k})\ket ,
\end{equation}
from which we obtain
\begin{equation}\label{eq:avg_green_fourier}
   \bra \widetilde{G}(\bm{k})\ket=\frac{1}{G_0(\bm{k})^{-1}-\widetilde{\Sigma}(\bm{k})}
      =\frac{1}{k^2-k_0^2-\widetilde{\Sigma}(\bm{k})},
\end{equation}
where $k=|\bm{k}|$.  The weak-extinction limit is defined by the condition $|\widetilde{\Sigma}(\bm{k})| \ll k_0^2$. Since
$ \bra \widetilde{G}(\bm{k})\ket$ is peaked around $k=k_0$, we can evaluate the self-energy $\Sigma(\bm{k})$ on-shell
for $k=k_0$. Making use of the identity
\begin{equation}
   \lim_{\epsilon\to 0^+}\frac{1}{x-x_0+i\epsilon}=\operatorname{PV}\left[\frac{1}{x-x_0}\right]-i\pi\delta(x-x_0),
\end{equation}
and taking the large-scale limit $\bm{q} \to 0$, it can be shown that
\begin{multline}\label{eq:prod_avg_green_fourier}
   \bra \widetilde{G}\left(\bm{k}+\frac{\bm{q}}{2}\right)\ket
   \bra \widetilde{G}^*\left(\bm{k}-\frac{\bm{q}}{2}\right)\ket
\\
      \sim
      -\frac{i\pi\delta[k^2-k_0^2-\re\widetilde{\Sigma}(k_0)]}{\bm{k}\cdot\bm{q}-i\im\widetilde{\Sigma}(k_0)}.
\end{multline}
The delta function fixes $k=k_r$, with $k_r^2=k_0^2+\re\widetilde{\Sigma}(k_0)$.
Next we define the specific intensity $\mathcal{I}(\bm{q},\bm{u})$ by
\begin{equation} \label{eq:def_specific_intensity}
   \frac{16\pi^3\omega}{\epsilon_0c^2k_r^3}\delta(k-k_r)\mathcal{I}(\bm{q},\bm{u})=
      \bra E\left(\bm{k}+\frac{\bm{q}}{2}\right)E^*\left(\bm{k}-\frac{\bm{q}}{2}\right)\ket ,
\end{equation}
where $\bm{u}=\bm{k}/k$ is a unit vector.
Inserting \eqs{eq:prod_avg_green_fourier} and (\ref{eq:def_specific_intensity}) into \eq{eq:bethe-salpeter_fourier}, performing the large-scale
limit $\bm{q} \to 0$ in the intensity vertex, and Fourier transforming with respect to $\bm{q}$, leads to the RTE
\begin{multline}\label{eq:RTE}
\bm{u}\cdot\nabla_{\bm{r}} \mathcal{I}(\bm{r},\bm{u}) + \frac{1}{\ell_e} \mathcal{I}(\bm{r},\bm{u}) = 
\frac{1}{\ell_s} \int p(\bm{u},\bm{u}') \mathcal{I}(\bm{r},\bm{u}') \, \ud\bm{u}' ,
\end{multline}
where $\ud\bm{u}'$ means integration over the unit sphere.  Here the extinction mean-free path $\ell_e$, the scattering
mean-free path $\ell_s$ and the phase function $p(\bm{u},\bm{u}')$ are defined by
\begin{align}\label{eq:extinction_length}
   \frac{1}{\ell_e} & = \frac{\im\widetilde{\Sigma}(k_0)}{k_r},
\\\label{eq:scattering_length}
   \frac{1}{\ell_s} & = \frac{1}{16\pi^2}\int\widetilde{\Gamma}(k_r\bm{u}',k_r\bm{u},k_r\bm{u}',k_r\bm{u})\ud\bm{u}',
\\\label{eq:phase_function}
  p(\bm{u},\bm{u}') & =  \frac{\ell_s}{16\pi^2} \widetilde{\Gamma}(k_r\bm{u}',k_r\bm{u},k_r\bm{u}',k_r\bm{u}) .
\end{align}
The specific intensity $\mathcal{I}(\bm{r},\bm{u})$ in \eq{eq:RTE} can be understood as a local and directional energy
flux, in agreement with the usual picture in radiative transfer theory~\cite{CHANDRASEKHAR-1950}. The regime in which
the transport of the average intensity is described by the RTE will be denoted hereafter by radiative transfer limit.

In the radiative transfer limit, making use of the weak-extinction and large scale ($\bm{q} \to 0$) approximations, it
is easy to show that \eq{eq:avg_abs_power_fourier} becomes
\begin{multline}\label{eq:avg_abs_power_kubo}
   \bra\mathcal{P}_a(\bm{r})\ket= \frac{1}{\ell_e} \int \mathcal{I}(\bm{r},\bm{u})\ud\bm{u} - \frac{1}{\ell_s} \int \mathcal{I}(\bm{r},\bm{u})\ud\bm{u} .
\end{multline}
Defining the average energy density $U$ by~\cite{CHANDRASEKHAR-1950}
\begin{align} \label{eq:avg_energy_density}
   U(\bm{r}) = \frac{1}{v_E}\int \mathcal{I}(\bm{r},\bm{u})\ud\bm{u} ,
\end{align}
where $v_E$ is the energy (or transport) velocity, and introducing the absorption mean free path $\ell_a$ such that
$\ell_a^{-1}=\ell_e^{-1}-\ell_s^{-1}$, we obtain
\begin{equation}\label{eq:avg_abs_power_transport}
   \bra\mathcal{P}_a(\bm{r})\ket=\frac{v_E}{\ell_a}U(\bm{r}) ,
\end{equation}
which is the usual expression of the average absorbed power density in radiative transfer
theory~\cite{CHANDRASEKHAR-1950}. We conclude that the (exact) expressions (\ref{eq:avg_abs_power}) and
(\ref{eq:avg_abs_power_fourier}) of the average absorbed power density, derived from multiple scattering theory, are
consitent with radiative transfer theory in the appropriate limit.

\section{Maximizing the average absorbed power}\label{sec:max}

In this section, we examine the possibility to maximize the average absorbed power in a disordered medium using
structural correlations.  As a model system, we consider a medium made of discrete absorbing scatterers, randomly
distributed in a transparent background with a predefined number density, the only degree of freedom being the positions
of the individual scatterers.

\subsection{A useful splitting of the average absorbed power}

Here we derive a splitting of the average absorbed power previously introduced in Ref.~\onlinecite{BIGOURDAN-2019}, 
that proved to be useful as a starting point for the maximization of absorption. 
We start by recalling the energy conservation law in a scattering medium. 
For scalar waves, we define the extinguished, absorbed and scattered power densities by
 \begin{align}\label{eq:ext_power}
   \mathcal{P}_e(\bm{r}) & = \frac{1}{2}\re\left[j(\bm{r})E_0^*(\bm{r})\right],
\\\label{eq:abs_power}
   \mathcal{P}_a(\bm{r}) & = \frac{1}{2}\re\left[j(\bm{r})E^*(\bm{r})\right],
\\\label{eq:sca_power}
   \mathcal{P}_s(\bm{r}) & = \bm{\nabla} \cdot \bm{J_s}(\bm{r}) ,
\end{align}
where $\bm{J}_s = 1/(2\omega\mu_0)\im\left[E_s(\bm{r})\bm{\nabla}E_s^*(\bm{r})\right]$ is the energy current of the
scattered field, $\mu_0$ being the vacuum permeability.  Energy conservation states that the extinguished power (the
power transferred from the incident field to the medium) is either absorbed or scattered, which reads as
\begin{equation}\label{eq:ext_abs_sca_power}
   \mathcal{P}_e(\bm{r})=\mathcal{P}_a(\bm{r})+\mathcal{P}_s(\bm{r}) .
\end{equation}
This local energy conservation law is derived in App.~\ref{app:conservation}.

In a disordered medium described statistically, all fields and induced sources are random variables, that can be written as the sum
of an average value and a fluctuation, in the form
\begin{equation}
   \left\{\begin{aligned}
      j & =\bra j\ket+\delta j,
   \\
      E & =\bra E\ket+\delta E,
   \\
      E_s & =\bra E_s\ket+\delta E_s,
   \end{aligned}\right.
\end{equation}
with all fluctuating terms averaging to zero.
Average quadratic quantities, such as the extinguished, absorbed and scattered power densities, can be cast in the form
\begin{equation}\label{eq:ext_abs_sca_power_average}
   \bra\mathcal{P}_{e,a,s}(\bm{r})\ket=\overline{\mathcal{P}}_{e,a,s}(\bm{r})+\widetilde{\mathcal{P}}_{e,a,s}(\bm{r}) ,
\end{equation}
where $\overline{\mathcal{P}}_{e,a,s}$ only depends on average quantities $\bra j\ket$, $\bra E\ket$
and $\bra E_s\ket$, and $\widetilde{\mathcal{P}}_{e,a,s}$ only depends on their fluctuating counterparts $\delta j$, $\delta E$ and
$\delta E_s$. Since the incident field $E_0$ is deterministic, $\bra E_0\ket=E_0$ and $\delta E_0=0$, which leads to
$\bra\mathcal{P}_e(\bm{r})\ket=\overline{\mathcal{P}}_e(\bm{r})$. Averaging \eq{eq:ext_abs_sca_power},
and making use of the above splitting, leads to
\begin{equation}\label{eq:energy_balance_full}
   \overline{\mathcal{P}}_e(\bm{r})=\overline{\mathcal{P}}_a(\bm{r})+\widetilde{\mathcal{P}}_a(\bm{r}) + \overline{\mathcal{P}}_s(\bm{r})
                                   +\widetilde{\mathcal{P}}_s(\bm{r}).
\end{equation}

The average field $\bra E\ket$ obeys the Dyson equation (\ref{eq:dyson_field}), which shows that it propagates in an
effective medium with properties defined by the self energy $\Sigma$. More precisely, in a statistically homogeneous and
isotropic medium, and in absence of non-locality (meaning that field variations on the scale of the correlation length
of the medium are disregarded), the average scattered field obeys an Helmholtz equation similar to
\eq{eq:wave_equation_sca}, with $\delta\epsilon$ replaced by an effective dielectric function, and $E$ replaced by $\bra
E \ket$ in the source term~\cite{SHENG-2006}.  Starting from this equation, and following the same steps as those
leading to \eq{eq:ext_abs_sca_power} (given in App.~\ref{app:conservation}), we obtain
\begin{equation}\label{eq:energy_balance_eff}
   \overline{\mathcal{P}}_e(\bm{r})=\overline{\mathcal{P}}_a(\bm{r})
                                   +\overline{\mathcal{P}}_s(\bm{r}).
\end{equation}
From \eqs{eq:energy_balance_full} and (\ref{eq:energy_balance_eff}),we immediately find that
\begin{equation}
   \widetilde{\mathcal{P}}_a(\bm{r})+\widetilde{\mathcal{P}}_s(\bm{r})=0,
\end{equation}
which also implies that
\begin{equation}\label{eq:avg_abs_power_bis}
   \bra\mathcal{P}_a(\bm{r})\ket=\overline{\mathcal{P}}_a(\bm{r})-\widetilde{\mathcal{P}}_s(\bm{r}).
\end{equation}
The first term on the right-hand side is by definition
\begin{equation}
   \overline{\mathcal{P}}_a(\bm{r})=\frac{1}{2}\re\left[\bra j(\bm{r})\ket\bra E^*(\bm{r})\ket\right] .
\end{equation}
It actually corresponds to the factorized contribution to $\bra\mathcal{P}_a(\bm{r})\ket_1$ in
\eq{eq:avg_abs_power_1}, namely
\begin{equation} \label{eq:Pa1_factorized}
   \overline{\mathcal{P}}_a(\bm{r})=
      \frac{\epsilon_0 c^2}{2\omega}\im\left[
         \int\Sigma(\bm{r},\bm{r}')\bra E(\bm{r}')\ket\bra E^*(\bm{r})\ket\ud^3r'
      \right].
\end{equation}
In the diagrammartic representation, this term results from all possible scattering sequences for $E$ and $E^*$ that are 
disconnected from one another. In the radiative transfer limit, it also corresponds to 
\begin{equation} \label{eq:Pa1_factorized_RTE}
   \overline{\mathcal{P}}_a(\bm{r})=\frac{v_E}{\ell_e}\overline{U}(\bm{r}) ,
\end{equation}
where $\overline{U}(\bm{r})$ is the ballistic energy density (\ie, the energy density associated to the average field).
Note that \eq{eq:Pa1_factorized_RTE} is deduced from \eq{eq:Pa1_factorized} by following the steps leading from
\eq{eq:avg_abs_power_fourier} to \eq{eq:avg_abs_power_kubo} (the full derivation is not written for the sake of brevity).
In the expression above, we clearly see that $\overline{\mathcal{P}}_a$ is not the power absorbed by the ballistic beam, which would
be proportional to $\ell_a^{-1}\overline{U}$. 

The second term on the right-hand side in \eq{eq:avg_abs_power_bis} is deduced by identification with the other terms in
\eqs{eq:avg_abs_power_1} and~\eqref{eq:avg_abs_power_2}, which gives
\begin{multline}\label{eq:fluc_sca_power_bis}
   \widetilde{\mathcal{P}}_s(\bm{r})=
      -\frac{\epsilon_0 c^2}{2\omega}\im\left[
         \int\Sigma(\bm{r},\bm{r}')\bra E(\bm{r}') E^*(\bm{r})\ket_c\ud^3r'
      \right.
\\\left.
        +\int\bra G^*(\bm{r}-\bm{\uprho},\omega)\ket
         \Gamma(\bm{r},\bm{\uprho},\bm{r}',\bm{\uprho}',\omega)
      \right.
\\\left.\vphantom{\int}\times
         \bra E(\bm{r}',\omega)E^*(\bm{\uprho}',\omega)\ket\ud^3r'\ud^3\rho\ud^3\rho'
      \right] ,
\end{multline}
where $\bra X X' \ket_c = \bra X X' \ket - \bra X \ket \bra X' \ket$ denotes the connected part of a correlation function.
In the radiative transfer limit, this corresponds to
\begin{align}
   \widetilde{\mathcal{P}}_s(\bm{r}) & =
      -\frac{v_E}{\ell_e}\widetilde{U}(\bm{r})+\frac{v_E}{\ell_s}U(\bm{r})
\\
   & = 
      -\frac{v_E}{\ell_a}\widetilde{U}(\bm{r})+\frac{v_E}{\ell_s}\overline{U}(\bm{r}) ,
\end{align}
where $\widetilde{U} = U - \overline{U}$ is the diffuse energy density. 
The total absorbed power $P_a$ is obtained by integrating the power density over the volume of the disordered medium. 
From \eq{eq:avg_abs_power_bis}, we immediately find that
\begin{equation} \label{eq:total_power_splitting}
\bra P_a\ket  =\overline{P}_a-\widetilde{P}_s .
\end{equation}
The term $\widetilde{P}_s$, which represents the scattered power of the fluctuating part of the field, is the integral
of $\widetilde{\mathcal{P}}_s(\bm{r})$ defined in \eq{eq:ext_abs_sca_power_average}, and can be
easily shown to be the integral of $|\delta E_s|^2$ over a sphere with radius tending to infinity and embedding the
disordered medium. This proves that $\widetilde{P}_s \geq 0$, thus suggesting a strategy to maximize the average
absorbed power by minimizing $\widetilde{P}_s$ and maximizing $\overline{P}_a$. This idea was previously examined in
Ref.~\cite{BIGOURDAN-2019} using numerical simulations, and will be analyzed theoretically below.  It is also useful to
note that the two terms on the right-hand side in \eq{eq:total_power_splitting} have the following expressions in the
radiative transfer limit:
\begin{align}
   \label{eq:avg_abs_power_bis_kubo}
   \overline{P}_a & = \frac{v_E}{\ell_e}\int\overline{U}(\bm{r})\ud^3r,
\\\label{eq:fluc_sca_power_bis_kubo}
   \widetilde{P}_s & = \frac{v_E}{\ell_s}\int\overline{U}(\bm{r})\ud^3r
                      -\frac{v_E}{\ell_a}\int\widetilde{U}(\bm{r})\ud^3r.
\end{align}
The structure of \eq{eq:fluc_sca_power_bis_kubo} confirms the positivity of $\widetilde{P}_s$. Indeed the first term
on the right-hand side corresponds to the power lost by the ballistic intensity due to scattering, which
contributes as a source for the diffuse intensity. The second term corresponds to absorption of the diffuse intensity, which
needs to be subtracted to count the net contribution of the diffuse intensity to the scattered power. 

\subsection{Minimizing $\widetilde{P}_s$}\label{subsec:min_p_s_wilde}

Minimizing $\widetilde{P}_s$ amounts to maximizing the scattering mean-free path $\ell_s$. Indeed, $\widetilde{P}_s$ 
is expected to vanish in the limit $\ell_s\to\infty$, as can be seen from \eq{eq:fluc_sca_power_bis_kubo}. The first
term on the right-hand side vanishes due to the prefactor $1/\ell_s$, and the second term vanishes since
$\widetilde{U}=0$ in absence of a source term generating a diffuse intensity. 

In order to maximize $\ell_s$, we now move to the particular case of absorbing particles in a transparent matrix.
We are thus left with the problem of connecting $\ell_s$ to structural correlations in a medium made of
discrete scatterers in a transparent background. In a dilute medium, the expansion of the intensity vertex $\Gamma$ can
be limited to the first diagram in \eq{eq:gamma_diag} (diagrams with increasing number of scattering events can be shown
to be proportional to increasing powers of the number density of scatterers $\rho$).  The first diagram corresponds to
\begin{equation}
   \Gamma(\bm{r},\bm{\uprho},\bm{r}',\bm{\uprho}') = k_0^4 \,
     C(\bm{r}-\bm{\uprho})\delta(\bm{r}-\bm{r}')\delta(\bm{\uprho}-\bm{\uprho}') ,
\end{equation}
where $C(\bm{R})$ is the correlation function defined in \eq{eq:correlation_epsilon}.
Inserting this expression into \eq{eq:scattering_length}, we obtain
\begin{equation} \label{eq:MFP_correlations}
   \frac{1}{\ell_s}=\frac{k_0^4}{16\pi^2}\int \widetilde{C}[k_r(\bm{u}'-\bm{u})]\ud\bm{u}' ,
\end{equation}
where $\widetilde{C}(\bm{q})$ is the Fourier transform of $C(\bm{R})$.
For a set of identical (non-resonant) absorbing scatterers occupying a volume $\mathcal{V} \to \infty$, 
the dielectric function can be written
\begin{multline}
   \epsilon(\bm{r})=1+\Delta\epsilon\sum_j \Theta(\bm{r}-\bm{r}_j)
\\
   \text{where}\quad
   \Theta(\bm{r}-\bm{r}_j)=\begin{cases}
      1 & \text{if $\bm{r}$ is in particle $j$,}
   \\
      0 & \text{otherwise,}
   \end{cases}
\end{multline}
The correlation function $C(\bm{r}-\bm{\uprho})$ reads as
\begin{multline}\label{eq:corr_particles}
   C(\bm{r}-\bm{\uprho})=|\Delta\epsilon|^2\int \Theta(\bm{r}-\bm{r}_a)\Theta(\bm{\uprho}-\bm{\uprho}_a)
\\\times
      \bra\sum_{i,j}\delta(\bm{r}_a-\bm{r}_i)\delta(\bm{\uprho}_a-\bm{r}_j) \ket_c
      \ud^3r_a\ud^3\rho_a ,
\end{multline}
which, after Fourier transformation, becomes
\begin{multline}\label{eq:corr_particles_fourier}
   \widetilde{C}(\bm{q})=\frac{|\Delta\epsilon\Theta(\bm{q})|^2}{\mathcal{V}}\int\exp[-i\bm{q}\cdot(\bm{r}_a-\bm{\uprho}_a)]
\\\times
      \bra\sum_{i,j}\delta(\bm{r}_a-\bm{r}_i)\delta(\bm{\uprho}_a-\bm{r}_j) \ket_c
      \ud^3r_a\ud^3\rho_a .
\end{multline}
We now introduce the structure factor defined by
\begin{equation}\label{eq:structure_factor}
   S(\bm{q})=\frac{1}{N}\bra
      \left|\sum_i \exp[i\bm{q}\cdot\bm{r}_i]\right|^2
   \ket ,
\end{equation}
where $N\to\infty$ is the number of scatterers, which can also be cast in the form
\begin{multline} \label{eq:structure_factor_bis}
   S(\bm{q})=\frac{1}{N}\int \exp[-i\bm{q}\cdot(\bm{r}_a-\bm{\uprho}_a)]
\\\times
   \bra
      \sum_{i,j} \delta(\bm{r}_a-\bm{r}_i)\delta(\bm{\uprho}_a-\bm{r}_j)
   \ket\ud^3r_a\ud^3\rho_a.
\end{multline}
From \eqs{eq:MFP_correlations} , (\ref{eq:corr_particles_fourier}) and (\ref{eq:structure_factor_bis}), the 
scattering mean free path can be written
\begin{equation}\label{eq:scattering_length_structure_factor}
   \frac{1}{\ell_s}=\frac{\rho k_0^4|\Delta\epsilon|^2}{16\pi^2}
      \int |\Theta[k_r(\bm{u}'-\bm{u})]|^2 \widetilde{S}[k_r(\bm{u}'-\bm{u})]\ud\bm{u}' ,
\end{equation}
where $\rho=N/\mathcal{V}$ is the number density of scatterers. Here we have introduced a corrected structure factor 
$\widetilde{S}(\bm{q})$, defined by \eq{eq:structure_factor_bis} with the average value replaced by its connected 
part $\bra \ldots \ket_c$. $\widetilde{S}(\bm{q})$ actually corresponds to the structure factor corrected from its forward
contribution, see for example Ref.~\onlinecite[section 2.1]{TORQUATO-2018} or the Supplemental Information
of Ref.~\onlinecite{LESEUR-2016-1}. Also note that $\Theta$ in \eq{eq:scattering_length_structure_factor}
accounts for the size of an individual scatterer. Actually the form factor is defined as
$F(\bm{q})=k_0^4k_r^2|\Delta\epsilon|^2|\Theta(\bm{q})|^2/(16\pi^2)$.

From \eq{eq:scattering_length_structure_factor}, we see that the limit $\ell_s \to \infty$ is reached if the structure
factor $\widetilde{S}(\bm{q})$ vanishes in a neighborhood of $\bm{q}=0$ containing the integration domain.  This
condition on $\widetilde{S}(\bm{q})$ defines a stealth hyperuniform distribution of the scatterers in
space~\cite{TORQUATO-2003,TORQUATO-2018}. It is interesting to note that \eq{eq:scattering_length_structure_factor}
holds in the multiple scattering regime (only the condition of weak-extinction $k_0 \ell_e \gg 1$ has been assumed).
This means that if $K$ is the size of the domain in which $\widetilde{S}(\bm{q})=0$ around $\bm{q}=0$, $\ell_s$ can be
considered infinite provided that $2k_r<K$. In a non absorbing medium, this coincides with the transparency regime
originally discussed in Ref.~\cite{LESEUR-2016-1}. In an absorbing medium, this transparency condition has to be
understood as the condition minimizing the scattered power $\widetilde{P}_s$.  Finally, let us note that since
\eq{eq:scattering_length_structure_factor} results from a first-order perturbative analysis (only the first diagram in
the intensity vertex $\Gamma$ has been taken into account), corrections to the value $1/\ell_s = 0$ discused above are
expected, due to the contribution of higher-order diagrams in $\Gamma$. This means that $1/\ell_s$ can be made very
small in the multiple scattering regime, but not exactly zero. In practice, the ``transparency'' regime is reached when
the effective $\ell_s$ becomes much larger than the size of the disordered medium. Beyond hyperuniformity, which is
a particular case of correlated disorder, the dependence of the scattering mean-free path on correlations has been studied in
different contexts such as in scattering in biological tissues~\cite{MAURICE-1957}, condensed matter
physics~\cite{ZIMAN-1979,BARBER-2019}, or scattering from colloidal suspensions~\cite{MARET-1990}. However stealth
hyperuniformity has the particularity to lead to a strong increase of the scattering mean-free path
~\cite{LESEUR-2016,TORQUATO-2018}.

\subsection{Maximizing $\overline{P}_a$}

\begin{figure}[htb]
   \centering
   \psfrag{L}[c]{$L$}
   \includegraphics[width=0.5\linewidth]{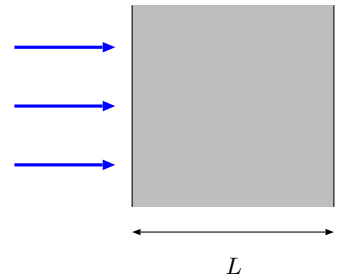}
   \caption{A slab of a scatterig medium with thickness $L$, illuminated by a plane-wave at normal incidence.}
   \label{fig:slab}
\end{figure}

Maximizing the average absorbed power $\bra P_a\ket$ in \eq{eq:total_power_splitting} also requires to maximize
 $\overline{P}_a$. Assuming that the condition $\widetilde{P}_s \simeq 0$ is satisfied, $\overline{P}_a$ can be obtained 
in the radiative transfer limit by solving the RTE \eq{eq:RTE} in absence of the source term on the right-hand side, and making
use of \eq{eq:avg_abs_power_bis_kubo}. For practical calculations, we consider a medium confined within a slab
with thickness $L$, illuminated by a plane wave at normal incidence, as represented schematically in Fig.~\ref{fig:slab}.

In this geometry, and for index matched media, we obtain
\begin{align}
   \overline{P}_a & =\frac{v_E S}{\ell_a}\int_0^L \frac{\mathcal{I}_0}{v_E}\exp\left[-\frac{s}{\ell_a}\right]\ud s
\\
   & =P_0\left[1-\exp\left(-\frac{L}{\ell_a}\right)\right] ,
   \label{eq:Pa_homogenized}
\end{align}
where we have assumed $\ell_e \simeq \ell_a$, and introduced $\mathcal{I}_0$ the specific intensity of the incident
wave, $S$ the transverse section of the slab (supposed to be tending to infinity), and $P_0=\mathcal{I}_0S$ the incident
power.  As expected, $\overline{P}_a$ reaches its maximum value  $\overline{P}_a = P_0$ for $L\gg\ell_a$. The condition
of index matching has been chosen for the sake of illustration. Index mismatch would result in multiplying the incident
power by a factor $1-R$, with $R$ being the reflectivity of the input interface, and by slightly increasing absorption
due to total internal reflection.  Keeping in mind the goal of designing efficient absorbers based on dilute materials,
these effects should remain weak compared to the absorption enhancement mechanism induced by structural correlations.

The fact that absorption reaches a maximum value when $\ell_a \ll L$ may seem obvious, since once scattering has been
suppressed, the medium is seen as homogenized by the propagating wave. Two subtle points need to be put forward. First,
the homogenization regime that is found is induced by changing the structure factor at a constant number density $\rho$.
In other words, it only results from spatial correlations in the positions of the scatterers (and not from a change in
the average distance between them). Second, although $\ell_s$ depends substantially on the structure factor
[\eq{eq:scattering_length_structure_factor}], the absorption mean free path $\ell_a$ is almost independent of structural
correlations. This interesting feature has been observed recently in numerical simulations based on an exact formulation
or a perturbative treatment of the scattering problem, in hard-sphere correlated systems~\cite{LESEUR-2016,WANG-2018} or
hyperuniform distributions~\cite{BIGOURDAN-2019}.  It is also supported by a more refined analysis based on pertubation
theory, which is presented in App.~\ref{app:abs_length_constant}. This means that even in the presence of structural
correlations, in the weak-extinction regime, the absorption mean-free path $\ell_a$ can be taken to be close to the
independent scattering (or Boltzmann) mean-free path $\ell_a^{B} = 1/(\rho \sigma_a)$, where $\sigma_a$ is the
absorption cross section of an individual scatterer. 

In summary, the scattered power $\widetilde{P}_s$ can be made arbitrarily small using a stealth hyperuniform
distribution of scatterers, by reaching the regime $\ell_s \gg L$. In this regime, the average absorption $\bra P_a\ket
\simeq \overline{P}_a$, and reaches a maximum provided that $\ell_a^{B} \ll L$, a condition that only depends on the
density $\rho$, and on the absorption cross section of individual scatterers. The fact that $\ell_s$ can be tuned
independently on $\ell_a^{B}$ makes structural correlations a practical degree of freedom in the maximization of the
average absorption. This result can also be understood
qualitatively  using a random walk picture, valid in the radiative transfer limit, in which the average absorbed power
can be written as
\begin{equation}
   \frac{\bra P_a\ket}{P_0}=1-\int_0^{\infty}P(s)\exp\left[-\frac{s}{\ell_a^B}\right]\ud s \, ,
\end{equation}
where $P(s)$ is the probability density of a path with length $s$. From this expression, we clearly see that $\bra
P_a\ket/P_0$ is maximized when the condition $\ell_a^B\ll L\ll\ell_s$ is fulfilled. In this case the integral
vanishes since $P(s)=0$ over the integral range limited by the exponential cutoff.

A medium made of absorbing and scattering particles in a transparent matrix has been chosen to provide a simple
model in which the substantial influence of structural correlations on absorption can be demonstrated. On the practical
side, disorder assemblies of scatterers in which absorption needs to be controlled are found, for example, in
soft-matter (a colloidal suspension of metallic nanoparticles, or a structural paint), in photonics (an ensemble of
quantum dots in a semiconductor structure), or in atomic physics (a cold atomic cloud). Given the nature of the
individual scatterers, maximizing (or more generally controlling) the collective absorption can be achieved using
structural correlations as the only available degree of freedom. For example, in structural paints, one uses small
absorbers to saturate the colors produce by scattering at specific wavelengths; in atomic physics and photonics, one can
use reduced scattering and enhanced absorption to favor collective effects.

\subsection{Critical optical thickness for absorption}\label{sect:critical_ba}

\begin{figure}[htb]
   \centering
   \psfrag{a}[c]{$b_a$}
   \psfrag{s}[cb]{$b_s$}
   \psfrag{B}[cb]{$b_{a,c}$}
   \includegraphics[width=0.7\linewidth]{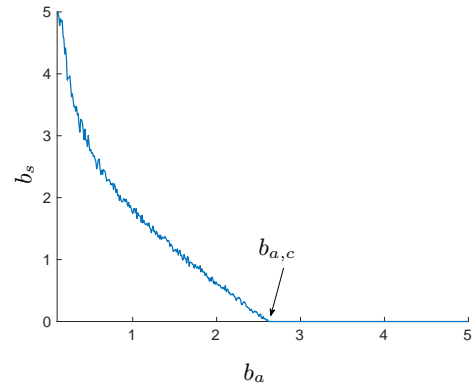}
   \caption{Scattering optical thickness $b_s$ giving the maximum average absorbed power versus the absorption thickness
   $b_a$. The calculation is performed in the radiative transfer limit, by solving the RTE using a Monte Carlo method
   for isotropic scattering \ie, a phase function $p(\bm{u},\bm{u}')=1/(4\pi)$ in \eq{eq:RTE}.  Beyond the critical absorption
   optical thickness $b_{a,c}$, the average absorbed power is maximum for $b_s=0$.  The value $b_{a,c} \simeq 2.61$ is
   found in this numerical simulation.}
   \label{fig:critical_b}
\end{figure}

Before concluding, it is worth analyzing the condition $\ell_a^{B} \ll L$ that maximizes $\overline{P}_a$ in more
details.  Introducing the scattering and absorption optical thicknesses $b_s=L/\ell_s$ and $b_a=L/\ell_a^{B}$,
respectively, we can expect that in the condition $b_s \simeq 0$ provided by stealth hyperuniformity, $\overline{P}_a$
reaches its maximum value for $b_a > b_{a,c}$, with $b_{a,c}$ a critical absorption optical thickness. Beyond $b_{a,c}$,
the homogenized medium can be considered as semi-infinite [which corresponds to $\overline{P}_a \simeq P_0$ in
\eq{eq:Pa_homogenized}], and the average absorbed power is maximized.  The value of $b_{a,c}$ can be determined
numerically. As an illustration, we find $b_{a,c}\sim 2.61$ in the radiative transfer limit, by solving the RTE using a
Monte Carlo method for isotropic scattering~\cite{NOEBAUER-2019}, as shown in~\fig{fig:critical_b}.  This value of
$b_{a,c}$ can also be determined using a semi-analytical approach, as described in App.~\ref{app:critical_b_a}.

\section{Conclusion}\label{sect:conclusion}

In conclusion, we have derived an expression for the average absorbed power in a disordered medium using multiple
scattering theory. This exact expression is consistent with radiative transfer theory in the weak-extinction regime.  In
this framework, we have discussed the possibility to maximize the average absorbed power $\bra P_a\ket$, using
structural correlations in the disorder as a degree of freedom. In a medium made of absorbing scatterers in a
transparent background, we have shown that a stealth hyperuniform distribution of the scatterers allows $\bra P_a\ket$
to reach its maximum value, provided that the absorption optical thickness is larger than a critical value.  We have
also shown that although the scattering mean-free path $\ell_s$ can be tuned using spatial correlations, the absorption
mean free path is almost independent of correlations, which is key point for a practical implementation of the proposed
strategy to maximize the absorbed power.  The analysis in this paper provides a clear theoretical support to the
observation made in Ref.~\onlinecite{BIGOURDAN-2019} based on numerical simulations. It provides a theoretical framework
for the design of blackbody-like non-resonant absorbers made of dilute materials, for broadband and omnidirectional
light, and for other kinds of waves. The techniques developed in this work could also serve as a basis to study the
relevant problem of minimization absorption in a scattering material. Another perspective is the study of absorption in
an scattering medium made of an absorbing background, which requires a substantially different model.

\section*{Acknowledgments}

This research is supported by LABEX WIFI (Laboratory of Excellence within the French Program ``Investments for the
Future'') under references ANR-10-LABX-24 and ANR-10-IDEX-0001-02 PSL*.

\appendix

\section{Ward identity}\label{app:Ward}

In the particular case of a non-absorbing medium (\ie $\im\delta\epsilon(\bm{r})=0$), it is well-known that the
self-energy $\Sigma$ and the intensity vertex $\Gamma$ are related by the Ward identity, as a consequence of energy
conservation. This fundamental relation is usually derived from the Bethe-Salpeter
equation~\cite{LAGENDIJK-1996,TSANG-2001-1,SHENG-2006,MONTAMBAUX-2007}. A Ward identity for the propagation of
electromagnetic waves in the presence of absorption has been derived in Ref.~\cite{LUBATSCH-2005}.  Here we check that
the result derived in Sec.~\ref{sec:avg_abs_power} is consistent with the usual Ward identity. To proceed, we
integrate \eq{eq:avg_abs_power_fourier} over the unbounded volume of a bulk and non-absorbing medium, which leads to
\begin{multline}\label{eq:to_ward}
   \im\left[\int \widetilde{\Sigma}\left(\bm{k}\right)\bra |E\left(\bm{k}\right)|^2\ket\frac{\ud^3k}{8\pi^3}
      +\int \bra\widetilde{G}^*\left(\bm{k}\right)\ket
   \right.
\\\left.\times
         \widetilde{\Gamma}\left(\bm{k},\bm{k},\bm{k}',\bm{k}'\right)
         \bra |E\left(\bm{k}'\right)|^2\ket\frac{\ud^3k}{8\pi^3}\frac{\ud^3k'}{8\pi^3}
      \right]=0.
\end{multline}
Making use of the reciprocity relations~\cite[Chap.~3, Problem~3.27]{APRESYAN-1996}
\begin{equation}
   \Gamma(\bm{r},\bm{\uprho},\bm{r}',\bm{\uprho}')=\Gamma^*(\bm{\uprho},\bm{r},\bm{\uprho}',\bm{r}')
      =\Gamma(\bm{r}',\bm{\uprho}',\bm{r},\bm{\uprho}),
\end{equation}
which leads to
\begin{equation}
   \widetilde{\Gamma}(\bm{k},\bm{k},\bm{k}',\bm{k}')=\widetilde{\Gamma}^*(\bm{k},\bm{k},\bm{k}',\bm{k}')
      =\widetilde{\Gamma}(\bm{k}',\bm{k}',\bm{k},\bm{k})
\end{equation}
in Fourier space, \eq{eq:to_ward} becomes
\begin{multline}\label{eq:to_ward_bis}
   \int \im\widetilde{\Sigma}\left(\bm{k}\right)\bra |E\left(\bm{k}\right)|^2\ket\frac{\ud^3k}{8\pi^3}
      =\int \im\bra\widetilde{G}\left(\bm{k}\right)\ket
\\\times
         \widetilde{\Gamma}\left(\bm{k},\bm{k},\bm{k}',\bm{k}'\right)
         \bra |E\left(\bm{k}'\right)|^2\ket\frac{\ud^3k}{8\pi^3}\frac{\ud^3k'}{8\pi^3} .
\end{multline}
This equation is consistent with the usual form of the Ward identity~\cite{LAGENDIJK-1996,TSANG-2001-1,SHENG-2006,MONTAMBAUX-2007}
\begin{equation}\label{eq:to_ward_bis}
   \im\widetilde{\Sigma}\left(\bm{k}\right)
      =\int \im\bra\widetilde{G}\left(\bm{k}'\right)\ket
         \widetilde{\Gamma}\left(\bm{k},\bm{k},\bm{k}',\bm{k}'\right)
         \frac{\ud^3k'}{8\pi^3}.
\end{equation}

\section{Energy balance in a scattering medium}\label{app:conservation}

In this appendix we derive the local energy balance \eq{eq:ext_abs_sca_power}.  We start by rewriting
\eq{eq:wave_equation_sca} in the form
\begin{equation}\label{eq:wave_equation_sca_bis}
   \Delta E_s +k_0^2E_s =-i\mu_0\omega j,
\end{equation}
with $j(\bm{r})=-i\omega \epsilon_0 \delta\epsilon(\bm{r})E(\bm{r})$ the induced current density in the medium.
From \eq{eq:wave_equation_sca_bis}, it is easy to show that
\begin{equation}
   E_s^*\Delta E_s- E_s \Delta E_s^* = -i\mu_0\omega j E_s^* - i\mu_0\omega j^* E_s,
\end{equation}
which can be rewritten as
\begin{align}\label{eq:conservation_energy}
   \bm{\nabla} \cdot (E_s^*\bm{\nabla} E_s- E_s \bm{\nabla} E_s^*) & = -2i\mu_0\omega \re[ j (E^*-E_0^*)].
\end{align}
Defining the energy current of the scattered field $\bm{J}_s$ by
\begin{equation}
   \bm{J}_s = \frac{1}{2\mu_0\omega} \im (E_s^*\bm{\nabla} E_s),
\end{equation}
\eq{eq:conservation_energy} can be written in the form of a conservation law
\begin{equation} \label{eq:local_conservation}
   \bm{\nabla} \cdot \bm{J}_s + \frac{1}{2} \re(j E^*)  = \frac{1}{2} \re(j E_0^*).
\end{equation}
In this equation, $\mathcal{P}_a=(1/2) \re(j E^*)$ is the absorbed power per unit volume, and $\mathcal{P}_s =
\bm{\nabla} \cdot \bm{J}_s$ can be understood as the scattered power per unit volume. The right-hand side $\mathcal{P}_e
= (1/2) \re(j E_0^*)$ is known as the extinguished power, and corresponds to the work done by the incident field on the
scattered medium. Equation (\ref{eq:local_conservation}) simply means that the power transferred from the incident field
$E_0$ to the medium ($\mathcal{P}_e$) is either absorbed ($\mathcal{P}_a$) or scattered ($\mathcal{P}_s$).  This
concludes the derivation of \eq{eq:ext_abs_sca_power}.

\section{Weak dependence of the absorption mean-free path on structural correlations}\label{app:abs_length_constant}

The weak dependence of the absorption mean-free path $\ell_a$ on structural correlations of disorder has been reported
in previous studies~\cite{LESEUR-2016,WANG-2018,BIGOURDAN-2019}. Here we show that this interesting feature can be
supported by a theoretical analysis based on a perturbative approach. We start with diagrammatic expansions of the
self-energy $\Sigma$ and the intensity vertex $\Gamma$ slightly different from that given in \eqs{eq:sigma_diag}
and~\eqref{eq:gamma_diag}, in which the free space Green functions $G_0$ are replaced by the average Green function
$\bra G \ket$ in all diagrams. Technically, this means that each diagram in the new expansions results from a partial
summation of an infinite number of the former diagrams. The new expansions for $\Sigma$ and $\Gamma$ take the form
\begin{multline}\label{eq:sigma_diag_bis}
   \Sigma = \underbrace{
      \begin{diagc}{4}
         \particule{2}
      \end{diagc}
   }_{\Sigma^{(1)}}+\underbrace{
      \begin{diagc}{10}
         \gmoy{2}{8}
         \correldeux{2}{8}
         \particule{2}
         \particule{8}
      \end{diagc}
   }_{\Sigma^{(2)}}+\underbrace{
      \begin{diagb}{22}
         \gmoy{2}{8}
         \gmoy{8}{14}
         \gmoy{14}{20}
         \correldeux{2}{14}
         \correldeux{8}{20}
         \particule{2}
         \particule{8}
         \particule{14}
         \particule{20}
      \end{diagb}
   }_{\Sigma^{(3)}}+\ldots
\end{multline}
and
\begin{multline}\label{eq:gamma_diag_bis}
   \Gamma = \underbrace{
   \begin{ddiagf}{4}
      \ccorreldeuxc{2}{-3}{2}{3}
      \pparticule{2}{3}
      \pparticule{2}{-3}
   \end{ddiagf}
   }_{\Gamma^{(2)}}+\underbrace{
   \begin{ddiagf}{16}
      \ccorreldeuxa{2}{14}
      \ccorreldeuxc{8}{-3}{8}{3}
      \ggmoy{2}{8}{3}
      \ggmoy{8}{14}{3}
      \pparticule{2}{3}
      \pparticule{8}{3}
      \pparticule{14}{3}
      \pparticule{8}{-3}
   \end{ddiagf}
   }_{\Gamma^{(3)\prime}}+\underbrace{
   \begin{ddiage}{16}
      \ccorreldeuxb{2}{14}
      \ccorreldeuxc{8}{-3}{8}{3}
      \ggmoy{2}{8}{-3}
      \ggmoy{8}{14}{-3}
      \pparticule{8}{3}
      \pparticule{2}{-3}
      \pparticule{8}{-3}
      \pparticule{14}{-3}
   \end{ddiage}
   }_{\Gamma^{(3)\prime\prime}}
\\
   +\underbrace{
   \begin{ddiag}{10}
      \ccorreldeuxc{2}{-3}{8}{3}
      \ccorreldeuxc{8}{-3}{2}{3}
      \ggmoy{2}{8}{3}
      \ggmoy{2}{8}{-3}
      \pparticule{2}{3}
      \pparticule{2}{-3}
      \pparticule{8}{3}
      \pparticule{8}{-3}
   \end{ddiag}
   }_{\Gamma^{(3)\prime\prime\prime}}+\ldots
\end{multline}
Here a thick solid line corresponds to the average Green function $\bra G\ket$. The equivalence between these expansions
and those used in Sec.~\ref{sec:avg_abs_power} can be verified by replacing each thick line with the iterative
solution of the Dyson equation for $\bra G \ket$.  For the sake of simplicity, we have represented only pair
correlations up to third-order scattering. These expansions are interesting since each diagram in $\Sigma$ can be
associated to its corresponding diagrams in $\Gamma$. Typically, $\Sigma^{(1)}$ will be considered alone, $\Sigma^{(2)}$
will be considered together with $\Gamma^{(2)}$, $\Sigma^{(3)}$ together with all $\Gamma^{(3)}$ diagrams, etc. Let us
start with first-order diagrams. In this case, only
\begin{equation}
   \widetilde{\Sigma}^{(1)}(\bm{r}-\bm{r}')=k_0^2\bra \delta\epsilon(\bm{r})\ket \delta(\bm{r}-\bm{r}')
\end{equation}
contributes in an infinite statistically homogeneous medium. The expression of the average absorbed power is simply
\begin{multline}
   \bra P_a^{(1)}\ket=\bra P_a^{(1)}\ket_1=
\\
   \frac{\epsilon_0c^2}{2\omega}\im\left[
      \int \Sigma^{(1)}(\bm{r},\bm{r}')\bra E(\bm{r}')E\*(\bm{r})\ket\ud^3r'\ud^3r
   \right]
\\
   =\frac{\epsilon_0\omega}{2}\bra \delta\epsilon''\ket \int\bra I(\bm{r})\ket\ud^3r ,
\end{multline}
where $\delta\epsilon''=\im\delta\epsilon$. As expected, $\bra P_a^{(1)}\ket=0$ in absence of absorption, that is, when
$\delta\epsilon''=0$. We also recall that the subscripts $1$ and $2$ are used here according to the definition in 
\eq{eq:avg_abs_power}. For the second-order contributions, we have to consider both
$\Sigma^{(2)}$ and $\Gamma^{(2)}$. Straightforward calculations lead to
\begin{multline}
   \bra P_a^{(2)}\ket=\bra P_a^{(2)}\ket_1+\bra P_a^{(2)}\ket_2=
\\
   \frac{\epsilon_0c^2}{4\omega}\im\left[
      \int k_0^4 \bra \delta\epsilon''(\bm{r})\delta\epsilon(\bm{r}')\ket_c \bra G(\bm{r}-\bm{r}')\ket
   \right.
\\\left.\vphantom{\int}\times
         \bra E(\bm{r}')E^*(\bm{r})\ket\ud^3r\ud^3r'
   \right].
\end{multline}
Here as well we find that $\bra P_a^{(2)}\ket$ vanishes in absence of absorption, confirming that all relevant diagrams
in $\Sigma$ and $\Gamma$ have been taken into account. Finally, for the third-order contributions, we need to consider
$\Sigma^{(3)}$ together with $\Gamma^{(3)\prime}$, $\Gamma^{(3)\prime\prime}$ and $\Gamma^{(3)\prime\prime\prime}$,
which leads to
\begin{widetext}
   \begin{multline}
      \bra P_a^{(3)}\ket=\bra P_a^{(3)}\ket_1+\bra P_a^{(3)}\ket_2=
      \frac{\epsilon_0c^2}{4\omega}\im\left[
            k_0^8\bra\delta\epsilon''(\bm{r})\delta\epsilon(\bm{r}''')\ket_c\bra\delta\epsilon(\bm{r}'')\delta\epsilon(\bm{r}')\ket_c
               \bra G(\bm{r}-\bm{r}'')\ket\bra G(\bm{r}''-\bm{r}''')\ket\bra G(\bm{r}'''-\bm{r}')\ket
      \right.
   \\\shoveright{\left.\times
            \bra E(\bm{r}')E^*(\bm{r})\ket\ud^3r\ud^3r'\ud^3r''\ud^3r'''
         +\int k_0^8\bra\delta\epsilon''(\bm{r}''')\delta\epsilon(\bm{r})\ket_c\bra\delta\epsilon^*(\bm{r}')\delta\epsilon(\bm{r}'')\ket_c
            \bra G^*(\bm{r}'''-\bm{r}')\ket
      \right.}
   \\\left.\vphantom{\int}\times
            \bra G(\bm{r}'''-\bm{r}'')\ket\bra G(\bm{r}''-\bm{r})\ket
            \bra E(\bm{r}')E^*(\bm{r})\ket\ud^3r\ud^3r'\ud^3r''\ud^3r'''
      \right] .
   \end{multline}
\end{widetext}
Again, we find that $\bra P_a^{(3)}\ket$ vanishes in absence of absorption, as it should be. For $i\in\{2,3\}$, we
clearly observe that the average absorbed power $\bra P_a^{(i)}\ket$ involves at least one correlation function of the
form $\bra \delta\epsilon''(\bm{r})\delta\epsilon(\bm{r}')\ket$. Conversely, considering the terms $\bra
P_a^{(i)}\ket_1$ and $\bra P_a^{(i)}\ket_2$ separately, we find that they only contain correlation functions involving
$\delta\epsilon$ and $\delta\epsilon^*$. Since, for most materials, the imaginary part of the dielectric function is
small compared to the real part, we conclude that $\bra \delta\epsilon''(\bm{r})\delta\epsilon(\bm{r}')\ket$ is small
compared to $\bra \delta\epsilon(\bm{r})\delta\epsilon(\bm{r}')\ket$ and $\bra
\delta\epsilon^*(\bm{r})\delta\epsilon(\bm{r}')\ket$. As a result, we must have
\begin{equation}
   \forall i>1, \bra P_a^{(i)}\ket\ll\left\{\bra P_a^{(i)}\ket_1,\bra P_a^{(i)}\ket_2\right\}.
\end{equation}
The only exception is for $i=1$ where
\begin{equation}
   \bra P_a^{(1)}\ket=\bra P_a^{(1)}\ket_1,
\end{equation}
a quantity that is not affected by structural correlations. Using this result in the expression of the average absorbed
power in the radiative transfer limit, we find that 
\begin{align} \nonumber
   \bra P_a\ket & =\frac{v_E}{\ell_a}\int U(\bm{r})\ud^3r=\sum_{i=1}^{\infty}\bra P_a^{(i)}\ket \sim \bra P_a^{(1)}\ket
\\ & \hphantom{=\frac{v_E}{\ell_a}\int U(\bm{r})\ud^3r}
   \sim \frac{v_E}{\ell_a^{B}}\int U(\bm{r})\ud^3r,
\end{align}
where $\ell_a^{B}$ is the independent scattering (or Boltzmann) absorption mean-free path, that would be obtained by
only considering the first-order approximation with $\Sigma^{(1)}$. In a set of uncorrelated discrete scatterers with number
density $\rho$, the independent scattering mean free path would be $\ell_a^{B}=1/(\rho\sigma_a)$, with $\sigma_a$ the
absorption cross-section of a single scatterer. From the result above, we can conclude that the absorption mean-free
path $\ell_a\simeq \ell_a^{B}$, and that the absorption mean free path weakly depends on structural correlations of
disorder.  Writing the two contribitions $\bra P_a\ket_1$ and $\bra P_a\ket_2$ in the same radiative transfer limit, we
obtain
\begin{align}\nonumber
   \bra P_a\ket_1 & =\frac{v_E}{\ell_e}\int U(\bm{r})\ud^3r=\sum_{i=1}^{\infty}\bra P_a^{(i)}\ket_1,
\\
   \bra P_a\ket_2 & =-\frac{v_E}{\ell_s}\int U(\bm{r})\ud^3r=\sum_{i=1}^{\infty}\bra P_a^{(i)}\ket_2 .
\end{align}
The series on the right-hand side cannot be simplified, showing that both $\ell_e$ and $\ell_s$ may substantially depend
on structural correlations, even in a regime in which $1/\ell_a = 1/\ell_e - 1/\ell_s$ is not affected.  Finally, we
point out that the analysis in this appendix is based on orders of magnitude, consistently with pertubative expansions,
and may fail in the presence of resonant scattering.

\section{Critical absorption optical thickness -- Analytical study}\label{app:critical_b_a}

\begin{figure}[htb]
   \centering
   \psfrag{L}[c]{$L$}
   \psfrag{R}[c]{$R$}
   \psfrag{T}[c]{$T$}
   \psfrag{s}[c]{$s$}
   \psfrag{sp}[c]{$s'$}
   \psfrag{t}[c]{$\theta$}
   \includegraphics[width=0.7\linewidth]{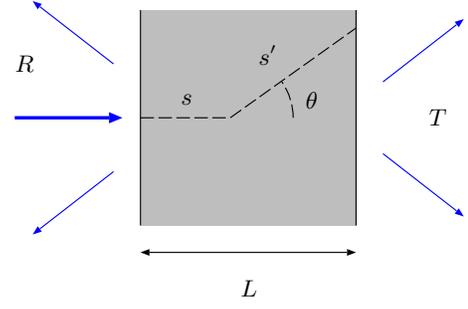}
   \caption{Geometry considered in the determination of the critical absorption thickness $b_{a,c}$. A slab of scattering
   and absorbing material is illuminated by a plane wave at normal incidence. $R$ ($T$) denotes the fraction of incident power
   that is reflected (transmitted).}
   \label{fig:system_bis}
\end{figure}

In this appendix we propose a semi-analytical approach to determine the critical absorption optical thickness introduced
in Sec.~\ref{sect:critical_ba}. With reference to Fig.~\ref{fig:system_bis}, we write the average absorbed power as
$\bra P_a\ket=P_0[1-(R+T)]$, where $P_0$ is the incident power, and $R$ and $T$ are the fractions of reflected and
transmitted power, respectively. $R$ and $T$ can be determined by solving the RTE~\eq{eq:RTE}.

Since the critical absorption thickness corresponds to the transition between $b_s=0$ and $b_s\ne 0$ (see
Fig.~\ref{fig:critical_b}), we can use a single-scattering approximation of the RTE. In this approximation, we can write
\begin{multline}
   R=\int_{s=0}^{\infty} \mu_s\exp(-\mu_ss) \int_{\mu=-1}^0 p(\mu,1) 
\\\times
   \int_{s'=0}^{\infty} \mu_s\exp(-\mu_ss')
   \exp\left[-\mu_a\left(s-\frac{s}{\mu}\right)\right]
\\\times
   \He[L-s]\He\left[s'+\frac{s}{\mu}\right]\ud s\ud\mu\ud s'
\end{multline}
where we use the notations $\mu_s=\ell_s^{-1}$, $\mu_a=\ell_a^{-1}$, $\mu_e=\mu_a+\mu_s$, $\mu=\cos\theta$, with the
angle $\theta$ defined in Fig.~\ref{fig:system_bis}, and $s$ denotes a path length. $\He$ is the Heaviside step
function, and $p(\mu,\mu')$ is the phase function integrated over the azimuthal angle $\varphi$. This way of writing the
solution to the RTE is common in slab geometries with azimuthal symmetry~\cite{CHANDRASEKHAR-1950}.  The first integral
over $s$ corresponds to ballistic propagation inside the medium, the second integral over $\mu$ corresponds to the
angular distribution of the single-scattering event, and the last integral over $s'$ describes backward propagation.
Similarly, for transmission we have
\begin{multline}
   T=\exp(-\mu_e L)+
      \int_{s=0}^{\infty} \mu_s\exp(-\mu_ss) \int_{\mu=-1}^0 p(\mu,1)
\\\times
      \int_{s'=0}^{\infty} \mu_s\exp(-\mu_ss')
      \exp\left[-\mu_a\left(s+\frac{L-s}{\mu}\right)\right]
\\\times
      \He[L-s]\He\left[s'-\frac{L-s}{\mu}\right]\ud s\ud\mu\ud s'.
\end{multline}
From the equations above we readily find that
\begin{multline} \label{eq:R_T_appendix}
   R+T=\exp(-\mu_e L)
      +\int_{\mu=0}^1 \frac{\mu_s}{\mu_e} \frac{\mu}{\mu+1}p(-\mu,1)
\\\times
      \left\{1-\exp\left[-\mu_eL\frac{\mu+1}{\mu}\right]\right\}\ud\mu
      +\int_{\mu=0}^1 \frac{\mu_s}{\mu_e}\frac{\mu}{\mu-1}p(\mu,1)
\\\times
      \left\{1-\exp\left[-\mu_eL\frac{\mu-1}{\mu}\right]\right\}
      \exp\left[-\frac{\mu_eL}{\mu}\right]\ud\mu.
\end{multline}
The critical absorption thickness $b_{a,c}$ is the solution to the implicit equation
\begin{equation}
   \left.\frac{\partial\bra P_a\ket}{\partial\mu_s}\right|_{\mu_s=0}=0,
\end{equation}
which, using \eq{eq:R_T_appendix}, becomes
\begin{multline}\label{eq:b_a}
   \int_0^1\frac{\mu p(-\mu,1)}{1+\mu}\left\{1-\exp\left[-\frac{b_{a,c}(1+\mu)}{\mu}\right]\right\}
\\
      +\frac{\mu p(\mu,1)}{1-\mu}\left\{\exp\left[-b_{a,c}\right]-\exp\left[-\frac{b_{a,c}}{\mu}\right]\right\}\ud\mu
\\
      -b_{a,c}\exp\left[-b_{a,c}\right]=0.
\end{multline}
This is the equation satsified by the critical absorption thickness. As an illustration, we find $b_{a,c}\simeq 2.61$
for an isotropic phase function [\ie $p(\bm{u},\bm{u}')=(4\pi)^{-1}$]. This result coincides with the full numerical
solution of the RTE presented in \fig{fig:critical_b}.

\end{document}